\documentclass[aps,floats]{revtex4}
\usepackage{amsmath,amssymb}
\usepackage{graphicx,epsfig}
\usepackage[greek,english]{babel}
\usepackage{bbold}
\begin{document}
\bibliographystyle {plain}

\pdfoutput=1
\def\oppropto{\mathop{\propto}} 
\def\opsimeq{\mathop{\simeq}}
\def\opoverderline{\mathop{\overline}}
\def\operarrow{\mathop{\longrightarrow}}
\def\opsim{\mathop{\sim}}

\def\fig#1#2{\includegraphics[height=#1]{#2}}
\def\figx#1#2{\includegraphics[width=#1]{#2}}

%\newcommand{\fig}[2]{\epsfxsize=#1\epsfbox{#2}} \reversemarginpar 

%%%%%%%%%%%%%%%%%%%%%%%%%%%%%%%%%%%%%%%%%%%%%%%%%%%%%%%%%%%%%%%%%%%%%%%%%%%%
\title{ microscopic Fluctuation Theory (mFT) for interacting Poisson processes } 

%%%%%%%%%%%%%%%%%%%%%%%%%%%%%%%%%%%%%%%%%%%%%%%%%%%%%%%%%%%%%%%%%%%%%%%%%%%%

\author{ C\'ecile Monthus }
 \affiliation{Institut de Physique Th\'{e}orique, 
Universit\'e Paris Saclay, CNRS, CEA,
91191 Gif-sur-Yvette, France}

\begin{abstract}

While the Macroscopic Fluctuation Theory (MFT) is a renormalized theory in the hydrodynamic limit based on a space-time local Lagrangian that is Gaussian with respect to the empirical current, C. Maes, K. Netocny and B. Wynants [Markov Proc. Rel. Fields. 14, 445 (2008)] have derived a microscopic Fluctuation Theory (mFT) for independent Markov jump processes based on a space-time local Lagrangian that is Poissonian with respect to the empirical flow, in direct relation with the general theory of large deviations at 'Level 2.5' for Markovian processes. Here we describe how this approach can be generalized to the presence of interactions, that can be either zero-range or involve neighbors, either for closed systems or for open systems with reservoirs.

\end{abstract}

\maketitle

\section{ Introduction }

Among the various approaches to describe non-equilibrium stochastic processes,
the idea that one should characterize the probabilities of dynamical trajectories
via the theory of large deviations (see the reviews \cite{oono,ellis,review_touchette} and references therein)
has been emphasized in various frameworks (see the reviews  \cite{derrida-lecture,harris_Schu,searles,harris,lazarescu_companion,lazarescu_generic}
and the PhD Theses \cite{vivien_thesis,chetrite_thesis,wynants_thesis,vroylandt_thesis,guioth_thesis} 
 or HDR Thesis masterpiece \cite{chetrite_HDR}).
For many-particles dynamics that are local in space and time,
the aim has been then to write the action for dynamical trajectories 
as an integral over space and time of an elementary space-time local Lagrangian
involving the local densities and flows of the conserved quantities. This goal has been achieved
on two different levels (macroscopic and microscopic) with different perspectives
 as we now recall.

\subsection{ Macroscopic Fluctuation Theory (MFT) for interacting driven diffusive systems}

\label{sec_mft}

The Macroscopic Fluctuation Theory (see the review \cite{mft} and references therein)
is a renormalized theory in the hydrodynamic limit where the measure over space-time trajectories is written as
\begin{eqnarray}
\int_{C_{x,t}} {\cal D} \rho(x,t) {\cal D} j(x,t) e^{ - \int dt \int dx {\cal L}^{Gauss} [ \rho(x,t) ,  j(x,t)] }
\label{mft}
\end{eqnarray}
The integral is over the empirical density $\rho(x,t) $ and the empirical current $ j(x,t) $ that are related by
the constraint $C_{x,t}$ of the empirical continuity equation for all $x$ and $t$
\begin{eqnarray}
C_{x,t} :   \ \ \partial_t \rho(x,t) + \nabla . j(x,t) =0
\label{continuity}
\end{eqnarray}
The Lagrangian that is local in space and time  follows the Gaussian form with respect to the empirical current $j$
\begin{eqnarray}
 {\cal L}^{Gauss} [ \rho ,  j] = [ j-J(\rho) ] \chi^{-1}(\rho)  [ j-J(\rho) ]
\label{gauss}
\end{eqnarray}
where the hydrodynamic current $J(\rho)$ contains a diffusive term with some diffusion coefficient $D(\rho)$
and a linear response term to the external field $E(x,t)$ that involves the mobility $\chi(\rho)$
\begin{eqnarray}
J(\rho) = - D(\rho) \nabla \rho + \chi(\rho) E
\label{jrho}
\end{eqnarray}
So the empirical density $\rho$ appears in the diffusion coefficient $D(\rho)$ and in the mobility $\chi(\rho)$
 that represent the constitutive properties of the system. 
The interactions between particles is encoded in the density-dependence of $D(\rho)$ and $\chi(\rho)$
as explained in detail in \cite{mft} with various examples of microscopic lattice models including exclusion processes
and zero-range processes.

\subsection{ microscopic Fluctuation Theory (mFT) for independent Markov Jump processes }

\label{sec_markovindep}

Since the standard classification of Large Deviations (see the reviews \cite{oono,ellis,review_touchette} and references therein)
involves three Levels, with Level 1 for empirical observables, 
Level 2 for the empirical density, and Level 3 for the empirical process,
the new name 'Level 2.5' has been introduced later as an intermediate level between Levels 2 and 3,
in order to refer to the joint properties of the empirical density and of the empirical flows.
For a single Markov process,  these large deviations at 'Level 2.5' for the joint distribution of the time-empirical-densities and the time-empirical-flows have been written in terms of explicit local-in-space rate functionals
within various frameworks, namely 
 for discrete-time and discrete-space Markov chains \cite{fortelle_thesis,fortelle_chain,review_touchette},
for continuous-time and discrete-space Markov jump processes \cite{fortelle_thesis,fortelle_jump,maes_canonical,maes_onandbeyond,wynants_thesis,chetrite_formal}
and for continuous-time and continuous-space diffusion processes \cite{wynants_thesis,maes_diffusion,chetrite_formal,engel}.
This  'Level 2.5' formulation allows to reconstruct any time-additive observable
of the dynamical trajectory via its decomposition in terms of the empirical densities and of the empirical flows,
and is thus closely related to the studies focusing on the generating functions of time-additive observables 
via deformed Markov operators that have attracted a lot of interest recently
in various models
 \cite{derrida-lecture,lecomte_chaotic,lecomte_thermo,lecomte_formalism,
lecomte_glass,kristina1,kristina2,chetrite_canonical,chetrite_conditioned,lazarescu_companion,lazarescu_generic,touchette_langevin,derrida-conditioned,bertin-conditioned}.

In Ref \cite{maes_onandbeyond}, this 'Level 2.5' for the joint distribution of the time-empirical-densities and the time-empirical-flows for a single Markov jump process has been extended to 'Level 2.5 in time' for
the joint distribution of the ensemble-empirical-occupations $N_t(x)$ and the ensemble-empirical-flows $q_t(y,x)$
for a large number of independent Markov jump processes involving the transitions rates $w_t(y,x)$ from site $x$ to site $y$
at time $t$. The output is the following measure on dynamical trajectories
\begin{eqnarray}
\int_{C_{x,t}} {\cal D} N_t(x) {\cal D} q_t(y,x) e^{ - \int dt \sum_x \sum_{y \ne x}  {\cal L}^{Poisson} [ N_t (.),  q_t(y,x)] }
\label{maes}
\end{eqnarray}
where we have slightly adapted the notations of Ref \cite{maes_onandbeyond} in order to 
make more obvious the comparison with the Macroscopic Fluctuation Theory described above.
The constraint $ C_{x,t}$ is the empirical dynamics that relates
the empirical occupation numbers $N_t(.)$ and the empirical flows $q_t[.,.]$
\begin{eqnarray}
C_{x,t} : \ \ \ \frac{d N_{t}(x)}{dt}  +  \sum_{y \ne x} \left( q_t(y,x)- q_t(x,y) \right) =0
\label{empiricaldyn}
\end{eqnarray}
This constraint thus represents the direct discrete-space analog of the continuity Eq. \ref{continuity}.
The space-time local Lagrangian follows the Poissonian form for the positive flows $q_t(y,x) \geq 0$
(instead of the Gaussian form with respect to the empirical current of Eq. \ref{gauss})
\begin{eqnarray}
 {\cal L}^{Poisson} [ N_t(.) , q_t(y,x) ] =  q_t(y,x)   \ln \left( \frac{ q_t(y,x)  }{ Q_t[y,x;N_t(.)]  } \right) - q_t(y,x) + Q_t[y,x;N_t(.)] 
\label{lpoisson}
\end{eqnarray}
Here $Q_t[y,x;N_t(.)] $ is the typical flow that the empirical occupations $N(.)$ are expected to produce,
i.e. the Lagrangian $ {\cal L}^{Poisson} [ N_t(.) , q_t(y,x) ]  $ vanishes if the empirical flow $q_t(y,x) $ takes this value $Q_t[y,x;N_t(.)] $.
So the typical flow $Q_t[y,x;N_t(.)]  $ is the analog 
of the hydrodynamic current $J(\rho) $ of Eq. \ref{jrho} within the Macroscopic Fluctuation Theory,
and represents the constitutive properties of the dynamics under study.
In particular, for independent particles that jump with some rates $w_t(y,x)  $, the 
 typical flow $Q_t[y,x;N_t(.)] $ is simply the product of this rate $w_t(y,x)  $ 
and of  the empirical number $N_t(x)$ of independent particles that are present on site $x$ at time $t$
\begin{eqnarray}
 Q^{indep}_t[y,x;N_t(.)] = w_t(y,x) N_t(x)
\label{Qindep}
\end{eqnarray}
This approach introduced in Ref \cite{maes_onandbeyond} for closed systems with a fixed number 
of independent Markov processes has been recently generalized to the case of open systems that can exchange particles with
reservoirs \cite{c_open}. Finally, the link with the so-called 'GENERIC' approach for non-equilibrium can be found in Refs \cite{generic1,generic2}
but is outside the scope of the present work.

\subsection{ Goal of the present paper : microscopic Fluctuation Theory (mFT) for interacting Poisson processes }

\label{sec_goal}

In the present paper, the goal is to generalize the microscopic Fluctuation Theory just described 
(with or without reservoirs) to
 the presence of interactions between the Markov Jump processes.
The particles will interact in the sense that the constitutive typical flow $Q_t[y,x;N_t(.)] $ from $x$ to $y$ at time $t$
produced by the occupation $N(.)$ will be different from the independent case of Eq. \ref{Qindep}
\begin{eqnarray}
 Q_t[y,x;N_t(.)] \ne    Q^{indep}_t[y,x;N_t(.)]= w_t(y,x) N_t(x)
\label{Qdep}
\end{eqnarray}

The simplest possible local interactions can be classified as follows :

(i) for Zero-Range Interactions, $Q_t[y,x;N_t(.)] $ only involves the occupation $N_t(x)$ of the initial site $x$ of the jump
\begin{eqnarray}
 Q^{ZR}_t[y,x;N_t(.)] =  Q_t[y,x;N_t(x)]
\label{Qzr}
\end{eqnarray}
The non-equilibrium properties of Zero-Range processes have attracted a lot of interest 
for the past fifty years (see the reviews \cite{zrp_Evans,zrp_EvansH} and references therein),
including recent works \cite{ozrp_levine,ozrp_harris,ozrp_popokov,ozrp_mukamel,ozrp_bertin}
on the open version with reservoirs at the boundaries.
For instance, Eq. \ref{Qzr} can follow a simple power-law with respect to the occupation $N_t(x)$
\begin{eqnarray}
 Q^{ZR}_t[y,x;N_t(.)] =  \lambda(x) [ N_t(x)]^g
\label{Qzrg}
\end{eqnarray}
With respect to the independent case of Eq. \ref{Qindep} that is linear with respect to $N_t(x)$,
the cases $g>1$ represent the systems where particles help each other to jump,
while the cases $0 \leq g<1$ represents the systems where particles hinder each other.

(ii) the next simplest case is when $Q_t[y,x;N_t(.)] $ only involves the occupations $N_t(x)$ and $N_t(y)$
of the initial site $x$ and of the final site $y$ of the jump

\begin{eqnarray}
 Q^{IF}_t[y,x;N_t(.)] =  Q_t[y,x;N_t(y);N_t(x)]
\label{Qif}
\end{eqnarray}

(iii) finally, the most general nearest-neighbor-interaction case would
be that $ Q_t[y,x;N_t(.)] $ depends not only on $N_t(x)$ and $N_t(y)$ but also on the occupations
of the nearest-neighbors of $x$ and $y$.

For these interacting microscopic models defined on the lattice,
the goal is thus to write the large deviations at level $2.5$
 for the empirical numbers of particles $N_t(x)$ and the empirical flows $q_t(y,x) $ defined on the lattice,
i.e. before the spatial coarse-graining analysis that leads to the Macroscopic Macroscopic Fluctuation Theory (MFT)
based on the Gaussian Lagrangian in continuous space, as recalled in section \ref{sec_mft}. 
Indeed, already for the well-known problem of the sum of a large number of random variables,
the continuous limit is always the Brownian motion, but 
this gaussian universal limit is only an expansion around the typical value,
while all the informations on the large deviation
properties of the initial problem, that depend on the details of the initial distribution,
 have been lost in this continuous limit \cite{review_touchette}.
Similarly for the case of independent Markov processes recalled in section \ref{sec_markovindep}, the Large deviations at level 2.5
have very different properties in discrete space (where the Poisson statistics of the jumps plays an essential role)
and in continuous space (where the Gaussian form for the empirical current $j$ is the only possibility), as stressed in particular in
\cite{c_ring} on the example of the behavior at large current.
As a consequence, for interacting microscopic models defined on the lattice,
it is natural to expect that some large deviation properties are not contained in the coarse-grained 
Gaussian Macroscopic Macroscopic Fluctuation Theory (MFT), but will require the 
large deviations at level $2.5$ for the empirical observables defined on the lattice.

\subsection{Organization of the paper }

The paper is organized as follows.
For interacting Poisson processes, the measure over dynamical trajectories is written
in terms of occupation $N_t(.)$ and flows $q_t(.,.)$ in section \ref{sec_flow},
and in terms of occupation $N_t(.)$ and currents $j_t(.,.)$ in section \ref{sec_current}.
The simplest possible application concerning directed one-dimensional interacting models between two reservoirs in 
described in section \ref{sec_directed}.
The generalization to other types of interacting Poisson processes is briefly discussed in section \ref{sec_compound}.
Our conclusions are summarized in \ref{sec_conclusion}.

\section{ Lagrangian in terms of occupation $N_t(.)$ and flows $q_t(.,.)$ }

\label{sec_flow}

\subsection{ Statistics of the elementary flow $q_t(y,x)  $ from $x$ to $y$ at time $t$ }

The probability of the elementary empirical flow $ q_t(y,x) $ from $x$ to $y$ at time $t$
that describes the fluctuations around the constitutive typical value $Q_t[y,x;N_t(.)] $
produced by the occupation $N(.)$,
involves some elementary Lagrangian $ {\cal L}_{Q_t[y,x;N_t(.)] } \left[ q_t(y,x) \right] $
\begin{eqnarray}
&&  {\cal P} \left[ q_t(y,x) \right] \propto e^{ - dt  {\cal L}_{Q_t[y,x;N_t(.)] } \left[ q_t(y,x) \right] }
\label{Plagrangian}
\end{eqnarray}
For a Poisson process, the increment $q_t(y,x)dt$ is one with probability $ Q_t[y,x;N_t(.)]dt$ and zero with probability $\left( 1-Q_t[y,x;N_t(.)]dt  \right) $,
so the generation function of parameter $\nu$ of this increment $q_t(y,x) dt  $ reads 
\begin{eqnarray}
< e^{ \nu q_t(y,x)dt } > &&= e^{ \nu } Q_t[y,x;N_t(.)]dt+ \left( 1-Q_t[y,x;N_t(.)]dt  \right)
 = e^{ \left( e^{ \nu }-1 \right) Q_t[y,x;N_t(.)] dt }
\label{gene}
\end{eqnarray}

In terms of the probability of Eq. \ref{Plagrangian},
the generating function of Eq. \ref{gene} corresponds to the integral
\begin{eqnarray}
< e^{ \nu q_t(y,x) dt } > &&= \int dq_t(y,x) {\cal P} \left[ q_t(y,x) \right] e^{ \nu q_t(y,x) dt }
=
 \int dq_t(y,x) e^{ dt \left[ \nu q_t(y,x) -   {\cal L}_{Q_t[y,x;N_t(.)] } \left[ q_t(y,x) \right]  \right] }
\label{geneinte}
\end{eqnarray}
The saddle-point evaluation yields that the result for the generating function of Eq. \ref{gene}
is the Legendre transform of ${\cal L}_{Q_t[y,x;N_t(.)] } \left[ q_t(y,x) \right]$, i.e. with simplified notations
\begin{eqnarray}
\nu q - {\cal L}_Q (q) && = (e^{ \nu } -1) Q 
\nonumber \\
\partial_q \left( \nu q -  {\cal L}_Q (q) \right) && =0
\label{legendre}
\end{eqnarray}
The inversion of this Legendre transform 
\begin{eqnarray}
 {\cal L}_Q (q) && = \nu q -  (e^{ \nu } -1) Q 
\nonumber \\
0 && = \partial_{\nu} \left( \nu q -  (e^{ \nu } -1) Q  \right) = q - e^{\nu} Q
\label{legendreinverse}
\end{eqnarray}
yields $ \nu(q) = \ln \frac{q}{Q}$ and
\begin{eqnarray}
 {\cal L}_Q (q) && =  q  \ln \left( \frac{q}{Q} \right) - q + Q 
\label{lagrangianq}
\end{eqnarray}
corresponding to the Poissonian form mentioned in Eq. \ref{lpoisson} of the introduction.

We should now discuss the validity of this calculation.
While Eq. \ref{gene} for the generating function can be considered as the definition of a Poisson process,
the result of Eq. \ref{lagrangianq} for the Lagrangian is obtained via the saddle-point evaluation of the integral
of Eq. \ref{geneinte}, which is expected to be a good approximation for sufficiently large $ Q_t[y,x;N_t(.)]$.
As a consequence, it is useful to rewrite the Lagrangian of Eq. \ref{lagrangianq} as
\begin{eqnarray}
 {\cal L}_Q (q) && =  Q I \left( \frac{q}{Q} \right) 
\nonumber \\
I(a)   && =   a \ln a - a+ 1 
\label{lagrangianqlargedev}
\end{eqnarray}
in order to make the link with large deviations language 
\cite{oono,ellis,review_touchette} : $Q$ is then the large parameter, while $I(a)$
is the rate function that describes the large deviations of the flow $q$ with respect to its typical value $Q$.
The validity of this condition can be thus summarized as follows :

(a) for the case of independent particles where the typical flow $Q^{indep}_t[y,x;N_t(.)] $ of  Eq. \ref{Qindep}
 is proportinal to the number $N_t(x)$ of particles at position $x$ at time $t$,
this means that the numbers $N_t(x)$ of particles on all sites $x$ at all times $t$ should remain large enough
in order to ensure the validity of the saddle-point evaluation described above,
as already stressed in \cite{maes_onandbeyond,c_open}.

(b) for the case of interacting particles, the condition is similarly 
 that the typical flow $Q^{dep}_t[y,x;N_t(.)]  $ of Eq. \ref{Qdep}
should remain large when the number of particles on all sites $x$ at all times $t$ remain large.
For instance, for the Zero-Range Process with the power-law dependence of Eq. \ref{Qzrg},
all positive powers $g>0$ correspond to a growth of $Q^{ZR}_t[y,x;N_t(.)]  $ with respect to $N_t(x)$
and are thus in the scope of the large deviation analysis above,
while the case $g=0$ corresponding to a finite typical flow $Q^{ZR}_t[y,x;N_t(.)]   $ is excluded.
Another important examples that are excluded 
are of course the exclusion processes where the numbers
of particles are limited to the two values $N_t(x) =0,1$ so that the typical flows $Q^{dep}_t[y,x;N_t(.)]   $ remain finite
and do not allow the saddle-point evaluation above.

\subsection{ Measure on dynamical trajectories during $[0,T]$ for closed systems }

For a given initial condition $N_{t=0}(x)$, 
the probability to observe an empirical dynamical trajectory on the time interval $[0,T]$
corresponding to the empirical occupations $N_{0 \leq t \leq T}(.)$ and 
to the empirical flows $q_{0 \leq t \leq T}(.,.)$ related by the empirical dynamics $C_{x,t}$ of Eq. \ref{empiricaldyn}
is obtained by summing the elementary contribution of Eq. \ref{lagrangianq} over the time $t \in [0,T]$ 
and over all the links $(y,x)$ of the system
\begin{eqnarray}
 \int_{C_{x,t}} {\cal D} N_t(x) {\cal D} q_t(y,x) e^{ -  \displaystyle 
\int_{0}^{T}dt 
\sum_{x } \sum_{y \ne x} \left[ q_t[y,x]  \ln \left( \frac{ q_t[y,x] }{ Q_t[y,x;N_t(.) ]} \right) - q_t[y,x]   +  Q_t[y,x;N_t(.)] \right]
  }
\label{ptrajmarkovjump}
\end{eqnarray}
It is exactly the same expression as in Eq. \ref{maes} for independent particles,
the only difference being in the behavior of the constitutive typical flows $Q_t[y,x;N_t(.) ] $
different from the independent case of Eq. \ref{Qindep} as explained in detail in section \ref{sec_goal}.

\subsection{  Generalization to open systems exchanging particles with reservoirs  }

\label{sec_open}

For independent particles, the generalization of the approach of \cite{maes_onandbeyond}
concerning closed systems to the case of open systems in contact with external reservoirs
has been described in detail in \cite{c_open}.
The formulation in the presence of interactions is straightforward
within the present framework and can be summarized as follows.

When particles can be exchanged between the system sites $x=1,2,..,\Omega$ 
and external reservoirs $r=1,..,R$,
the empirical dynamics of Eq. \ref{empiricaldyn} becomes for sites $x$ of the system
\begin{eqnarray}
C^R_{x,t} : \ \ \ \frac{d N_{t}(x)}{dt}  +  \sum_{y \ne x} \left( q_t(y,x)- q_t(x,y) \right) 
+\sum_{r} \left( q_t(r,x)- q_t(x,r) \right) =0
\label{empiricaldynr}
\end{eqnarray}
while the occupations $N_t(r)$ of reservoirs $r=1,..,R$ are not free dynamical variables but are
external control parameters that can be chosen to be fixed in time $N_t(r)=N(r)$ or to depend on time
depending the cases that one wishes to study (see more details and examples in \cite{c_open}).

One then needs to take into account the Lagrangian contributions concerning  the flows of particles 
between the system and the reservoirs.
Although other choices are of course possible,
we will consider as in \cite{c_open},
that the elementary incoming flow $q_t(y,r)$ from a given reservoir $r$ to a given site $y$ at time $t$
is a Poisson process of constitutive typical value $Q^{incoming}_t[y,r;N_t(.)]  $,
so that the corresponding Lagrangian follows the same Poissonian form as Eq. \ref{lpoisson}
\begin{eqnarray}
 {\cal L}^{incoming} [ N_t(.) , q_t(y,r) ] =  q_t(y,r)   \ln \left( \frac{ q_t(y,r)  }{ Q^{incoming}_t[y,r;N_t(.)]  } \right) - q_t(y,r) + Q^{incoming}_t[y,r;N_t(.)] 
\label{incoming}
\end{eqnarray}
Similarly, the elementary outgoing flow $q_t(r,x)$ from a given site $x$  to a given reservoir $r$ at time $t$
will be assumed to be a Poisson process of constitutive typical value $Q^{outgoing}_t[r,x;N_t(.)]  $,
with the corresponding Lagrangian
\begin{eqnarray}
 {\cal L}^{outgoing} [ N_t(.) , q_t(r,x) ] =  q_t(r,x)   \ln \left( \frac{ q_t(r,x)  }{ Q^{outgoing}_t[r,x;N_t(.)]  } \right) - q_t(r,x) + Q^{outgoing}_t[r,x;N_t(.)] 
\label{outgoing}
\end{eqnarray}

Then the measure of Eq. \ref{ptrajmarkovjump} for closed systems become for open systems
\begin{eqnarray}
 \int_{C^R_{x,t}} {\cal D} N_t(x) {\cal D} q_t(y,x) && e^{ -  \displaystyle 
\int_{0}^{T}dt 
\sum_{x } \sum_{y \ne x} \left[ q_t[y,x]  \ln \left( \frac{ q_t[y,x] }{ Q_t[y,x;N_t(.) ]} \right) - q_t[y,x]   +  Q_t[y,x;N_t(.)] \right]
  }
\nonumber \\
&& e^{ -  \displaystyle 
\int_{0}^{T}dt 
\sum_{y } \sum_{r} \left[  q_t(y,r)   \ln \left( \frac{ q_t(y,r)  }{ Q^{incoming}_t[y,r;N_t(.)]  } \right) - q_t(y,r) + Q^{incoming}_t[y,r;N_t(.)]  \right]
  }
\nonumber \\
&& e^{ -  \displaystyle 
\int_{0}^{T}dt 
\sum_{x } \sum_{r} \left[  q_t(r,x)   \ln \left( \frac{ q_t(r,x)  }{ Q^{outgoing}_t[r,x;N_t(.)]  } \right) - q_t(r,x) + Q^{outgoing}_t[r,x;N_t(.)]  \right]
  }
\label{ptrajmarkovjumpR}
\end{eqnarray}

So for each specific application, one can choose as one wishes the constitutive typical values $Q_t[y,x;N_t(.) ] $ of the bulk flows,
the constitutive typical values $  Q^{incoming}_t[y,r;N_t(.)] $ of the incoming flows, and the 
constitutive typical values $ Q^{outgoing}_t[r,x;N_t(.)]$ of the outgoing flows, in order to describe the interactions one is interested in.

\section{ Lagrangian in terms of occupation $N_t(.)$ and currents $j_t(.,.)$ }

\label{sec_current}

\subsection{  Statistics of the elementary current $j_t(y,x)  $ from $x$ to $y$ at time $t$  }

For a given pair of sites, one can choose some order $x < y$, 
and decide to replace the two flows $q_t(y,x)$ and $q_t(x,y)$ in the two directions of this link
by the antisymmetric and symmetric parts called respectively the current $ j_t(y,x) $ and the activity $a_t(y,x) $
\cite{maes_onandbeyond}
\begin{eqnarray}
j_t(y,x)  && =q_t(y,x) - q_t(x,y)= - j_t(x,y)
\nonumber \\
a_t(y,x) && = q_t(y,x) + q_t(x,y)  = a_t(x,y)
\label{jafromq}
\end{eqnarray}

When one is not interested in the activity $a_t(y,x)$,
one wishes to write the probability to observe the empirical current
$j_t(y,x)  $ alone with some elementary Lagrangian $ {\cal L}_{Q_t[y,x;N_t(.)],Q_t[x,y;N_t(.) } \left[ j_t(y,x) \right]$
\begin{eqnarray}
  {\cal P} \left[ j_t(y,x) \right] \propto e^{ - dt  {\cal L}_{Q_t[y,x;N_t(.)],Q_t[x,y;N_t(.) } \left[ j_t(y,x) \right] }
\label{Plagrangianj}
\end{eqnarray}
It is instructive to consider two methods to derive the form of this Lagrangian as we now describe.

\subsubsection{ Derivation via optimization over the activity \cite{maes_onandbeyond}   }

With simplified notations, the Lagrangian for the two flows $q_t(y,x)=q_+$ and $q_t(x,y)=q_-$ on the same link $x<y$
is simply the sum of the two Poissonian lagrangians of Eq. \ref{lagrangianq}
\begin{eqnarray}
{\cal L}_{Q_+,Q_-}(q_+,q_-) && =  {\cal L}_{Q_+} ( q_+ ) + {\cal L}_{Q_-} ( q_-)   =
 q_+  \ln \left( \frac{q_+}{Q_+} \right) - q_+ + Q_+
+  q_-  \ln \left( \frac{q_-}{Q_-} \right) - q_- + Q_-
\label{lagrangiansumlink}
\end{eqnarray}
Via the change of variable into the current $j$ and the activity $a$ (Eq. \ref{jafromq}),
this Lagrangian becomes
\begin{eqnarray}
{\cal L}_{Q_+,Q_-}(j,a) && =  {\cal L}_{Q_+} ( q_+= \frac{a+j}{2} ) + {\cal L}_{Q_-} ( q_-= \frac{a- j}{2} )   
\nonumber \\
&& =  \frac{a+j}{2}  \ln \left( \frac{a+j}{2 Q_+} \right) - \frac{a+j}{2} + Q_+
+  \frac{a- j}{2}  \ln \left( \frac{a-j}{2Q_-} \right) - \frac{a- j}{2}  + Q_-
\nonumber \\
&& =  \frac{a}{2}  \ln \left( \frac{a^2-j^2}{ 4 Q_+ Q_- } \right)
+ \frac{j}{2}  \ln \left( \frac{a+j}{a-j} \times \frac{Q_-}{Q_+}\right)
-a +Q_+ + Q_-
\label{lagrangianja}
\end{eqnarray}

The optimization with respect to the activity $a$
\begin{eqnarray}
0= \partial_a {\cal L}_{Q_+,Q_-}(j,a) 
&& =  \frac{1}{2}  \ln \left( \frac{a^2-j^2}{ 4 Q_+ Q_- } \right)
\label{lagrangianjaderi}
\end{eqnarray}
yields  the optimal activity
\begin{eqnarray}
 a_{opt}(j) = \sqrt{ j^2 + 4 Q_+ Q_- } 
\label{aopt}
\end{eqnarray}
that one plugs into Eq. \ref{lagrangianja} to obtain the Lagrangian of the current alone \cite{maes_onandbeyond}
\begin{eqnarray}
{\cal L}_{Q_+,Q_-}(j) && = {\cal L}_{Q_+,Q_-}(j, a_{opt}(j) )
=  \frac{a_{opt}(j) }{2}  \ln \left( \frac{a^2_{opt}(j) -j^2}{ 4 Q_+ Q_- } \right)
+ \frac{j}{2}  \ln \left( \frac{a_{opt}(j) +j}{a_{opt}(j) -j} \times \frac{Q_-}{Q_+}\right)
-a_{opt}(j)  +Q_+ + Q_-
\nonumber \\
&& =  j  \ln \left( \frac{j +\sqrt{ j^2 + 4 Q_+ Q_- } }{ 2 Q_+}\right)
-   \sqrt{ j^2 + 4 Q_+ Q_- } +Q_+ + Q_-
\label{ljopt}
\end{eqnarray}

\subsubsection{ Alternative derivation via generating functions }

Another way to understand how the form of Eq. \ref{ljopt} appears
consists in rewriting the generating function of the
 two Poissonian flows $ q_t(y,x) =q_+$ and $ q_t(x,y) =q_-$ (Eq \ref{gene})
\begin{eqnarray}
< e^{ \left[ \nu_+ q_+ + \nu_- q_- \right] dt } > &&=
 e^{ \left[ \left( e^{ \nu_+ }-1 \right) Q_+ +  \left( e^{ \nu_-} -1 \right) Q_-   \right]dt }
\label{genetwo}
\end{eqnarray}
 as a generating function for the current $ j $ and the activity $ a$ of Eq. \ref{jafromq}
\begin{eqnarray}
< e^{ \left[ \nu_+ q_+ + \nu_- q_- \right] dt } >   &&=
< e^{ \left[ \frac{\nu_+- \nu_- }{2} j + \frac{\nu_++ \nu_- }{2} a \right] dt } > 
\label{genetwobis}
\end{eqnarray}
In particular, the choice $\nu_- = - \nu_+$ yields that the generating function of the current $j $ alone reads
\begin{eqnarray}
< e^{ \nu j   dt } > &&=
 e^{ \left[ \left( e^{ \nu }-1 \right) Q_++  \left( e^{ - \nu }-1 \right) Q_-   \right]dt }
\label{genej}
\end{eqnarray}

In terms of 
the probability to observe the empirical current
$j  $ of Eq. \ref{Plagrangianj},
the generating function of Eq. \ref{genej} corresponds to the integral
\begin{eqnarray}
< e^{ \nu j dt } > &&= \int dj  e^{ dt \left[ \nu j -  
{\cal L}_{Q_+,Q_-} (j)  \right] }
\label{geneintej}
\end{eqnarray}
The saddle-point evaluation yields that the the generating function of Eq. \ref{genej}
is the Legendre transform of $ {\cal L}_{Q_+,Q_-} (j) $
\begin{eqnarray}
\nu j - {\cal L}_{Q_+,Q_-} (q) && = (e^{ \nu } -1) Q_+ + (e^{ -\nu } -1) Q_-
\nonumber \\
\partial_j \left( \nu j -  {\cal L}_{Q_+,Q_-} (j) \right) && =0
\label{legendrej}
\end{eqnarray}
The inversion of this Legendre transform 
\begin{eqnarray}
 {\cal L}_{Q_+,Q_-} (j) && = \nu j -  (e^{ \nu } -1) Q_+ - (e^{ -\nu } -1) Q_- 
\nonumber \\
0 && = \partial_{\nu} \left(   \nu j -  (e^{ \nu } -1) Q_+ - (e^{ -\nu } -1) Q_-   \right) = 
j - e^{ \nu } Q_+ + e^{ -\nu }  Q_- 
\label{legendreinversej}
\end{eqnarray}
yields 
\begin{eqnarray}
\nu(j)  && = \ln \left( \frac{j + \sqrt{ j^2+4 Q_+ Q_-} }{2 Q_+}  \right)
\label{nuj}
\end{eqnarray}
and 
\begin{eqnarray}
 {\cal L}_{Q_+,Q_-} (j) && =   j  \ln \left( \frac{j + \sqrt{ j^2+4 Q_+ Q_-} }{2 Q_+}  \right)
-  \sqrt{ j^2+4 Q_+ Q_-}  +Q_+ + Q_- 
\label{lagrangianj}
\end{eqnarray}
in agreement with Eq. \ref{ljopt}.

\subsubsection{ Properties of the Lagrangian for the current }

Returning to the complete notations, the probability of Eq. \ref{Plagrangianj}
 to observe the empirical current $j_t(y,x)  $ from $x$ to $y$ at time $t$ involves the Lagrangian
\begin{eqnarray}
 {\cal L}_{Q_t[y,x;N_t(.)],Q_t[x,y;N_t(.) } \left[ j_t(y,x) \right]
&& 
=
 j_t(y,x)  \ln \left( \frac{j_t(y,x) + \sqrt{ j^2_t(y,x)+4 Q_t[y,x;N_t(.)] Q_t[x,y;N_t(.)]} }{2 Q_t[y,x;N_t(.)]}  \right)
\nonumber \\
&& -  \sqrt{ j^2_t(y,x)+4 Q_t[y,x;N_t(.)] Q_t[x,y;N_t(.)]}  +Q_t[y,x;N_t(.)] + Q_t[x,y;N_t(.)] 
\label{lagrangianjfull}
\end{eqnarray}
which is thus very different from the Gaussian form of Eq. \ref{mft} of the macroscopic fluctuation theory
in the hydrodynamic limit.
A detailed discussion of the Gaussian approximation for small fluctuations around typical values
can be found in \cite{maes_onandbeyond}.

An important property of the form of Eq. \ref{lagrangianjfull}
is that
 the ratio of the probabilities to observe the current $j_t(y,x)$ and to observe the opposite current $(-j_t(y,x))$
simplifies into a linear term with respect to the current in the exponential 
\begin{eqnarray}
\frac{  {\cal P} \left[ j_t(y,x) \right] }{  {\cal P} \left[ -j_t(y,x) \right] }
\propto \frac{ e^{ - dt  {\cal L}_{Q_t[y,x;N_t(.)],Q_t[x,y;N_t(.) } \left[ j_t(y,x) \right] } }
{ e^{ - dt  {\cal L}_{Q_t[y,x;N_t(.)],Q_t[x,y;N_t(.) } \left[ - j_t(y,x) \right] } } 
= e^{ dt j_t(y,x) \ln \left( \frac{Q_t[y,x;N_t(.)]}{Q_t[x,y;N_t(.)]}  \right)  }
\label{symj}
\end{eqnarray}
As discussed in detail in Ref. \cite{maes_onandbeyond},
this is an example of the Gallavotti-Cohen symmetry 
\cite{derrida-lecture,harris_Schu,searles,harris,chetrite_thesis,lazarescu_companion,lazarescu_generic}
that can be interpreted in terms of the entropy production that measures the irreversibility.
The link with various decompositions of entropy contributions can be also found in Ref. \cite{maes_onandbeyond}.

\subsection{ Measure on dynamical trajectories during $[0,T]$ for closed systems }

The empirical dynamics $C_{x,t}$ of Eq. \ref{empiricaldyn}
can be rewritten in terms of the currents $j_t(y,x) $ only
\begin{eqnarray}
C_{x,t}     :   \ \ \    \frac{d N_{t}(x)}{dt}  +  \sum_{y \ne x}  j_t(y,x)  =0
\label{empiricaldynj}
\end{eqnarray}
For a given initial condition $N_{t=0}(x)$, 
the probability to observe an empirical dynamical trajectory on the time interval $[0,T]$
corresponding to the empirical occupation numbers $N_{0 \leq t \leq T}(.)$ and 
to the empirical currents $j_{0 \leq t \leq T}(.)$ related by the empirical dynamics of Eq. \ref{empiricaldynj}
is obtained by summing the elementary Lagrangian of Eq. \ref{lagrangianjfull} over the time $t \in [0,T]$ and 
over the ordered links $x<y$ of the system
\begin{eqnarray}
&& \int_{C_{x,t}} {\cal D} N_t(x) {\cal D} j_t(y,x) e^{ -   A [ N_{0 \leq t \leq T}(.) ; j_{0 \leq t \leq T}[.,.] ]    }
\label{ptrajmarkovjumpj}
\end{eqnarray}
with the action
\begin{eqnarray}
A [ N_{0 \leq t \leq T}(.) ; j_{0 \leq t \leq T}(.,.) ] 
&& =  -
\int_{0}^{T}dt 
\sum_{x < y} 
\Bigg( j_t(y,x)  \ln \left( \frac{j_t(y,x) + \sqrt{ j^2_t(y,x)+4 Q_t[y,x;N_t(.)] Q_t[x,y;N_t(.)]} }{2 Q_t[y,x;N_t(.)]}  \right)
\nonumber \\
&&-  \sqrt{ j^2_t(y,x)+4 Q_t[y,x;N_t(.)] Q_t[x,y;N_t(.)]}  +Q_t[y,x;N_t(.)] + Q_t[x,y;N_t(.)] \Bigg)
\label{actionj}
\end{eqnarray}
Again, the expression is the same as for independent particles \cite{maes_onandbeyond},
the only difference being in the behavior of the constitutive typical flows $Q_t[y,x;N_t(.) ] $
different from the independent case of Eq. \ref{Qindep} as discussed in section \ref{sec_goal}.

\subsection{ Measure on dynamical trajectories during $[0,T]$ for open systems with reservoirs }

Since we have already described the measure for open systems in terms of the flows $q_t(y,x)$
in section \ref{sec_open}, it seems enough to say here that one just needs to add the contributions
of the Lagrangians concerning the currents between the systems sites $x=1,2,.. \Omega$ and the reservoirs $r=1,2,..,R$.

\section{Application to open one-dimensional directed interacting models }

\label{sec_directed}

The simplest geometry where the general framework described above can be applied
is a one-dimensional lattice of $L$ sites between two reservoirs.
A further simplification consists in considering the directed version of interacting Poisson models,
where the flows occur only in the forward direction and not in the backward direction.

\subsection{ Models and notations }

In this section, we consider that the system contains $L$ sites $x=1,2,..,L$ between two reservoirs at $r=0$ and $r=L+1$.
As explained in section \ref{sec_open}, the occupations $N_t(x)$ for the system sites $x=1,2,..,L$
are dynamical variables, while the occupations $N_t(r=0)$ and $N_t(r=L+1)$ of the reservoirs $r=0$ 
and $r=L+1$ are two external control parameters.

Inside the system, the dynamics is defined by the constitutive
typical values $Q_t[x+1,x;N_t(.) ] $ of the bulk flows for $x=1,..,L-1$
that will be denoted by the simplified notation of the constitutive typical currents $J_t[x,N_t(.)]$ (since there is no backward flows)
\begin{eqnarray}
Q_t[x+1,x;N_t(.) ]  && \equiv  J_t[x,N_t(.)]
\label{QJ}
\end{eqnarray}

The constitutive typical values for incoming flow from the reservoir $r=0$ to the first site $x=1$ of the system
and of the outgoing flow from the last site $L$ of the system towards the reservoir $r=L+1$ 
will be also denoted by the simplified notations of constitutive typical currents
\begin{eqnarray}
Q^{incoming}_t[1,0;N_t(.) ]  && \equiv  J_t[0,N_t(.)]
\nonumber \\
Q^{outgoing}_t[L+1,L;N_t(.) ]  && \equiv  J_t[L,N_t(.)]
\label{Qout}
\end{eqnarray}

\subsection{ Measure on dynamical trajectories during $[0,T]$  }

The empirical flows will be denoted with the simplified notation of currents as in Eq. \ref{QJ}
(since again there is no backward flows)
\begin{eqnarray}
q_t(x+1,x)  && \equiv  j_t(x)
\label{qj}
\end{eqnarray}
For $x=1,2,..,L$ the empirical dynamics of Eq. \ref{empiricaldynr} reads
\begin{eqnarray}
C^R_{x,t} : \ \ \ \frac{d N_{t}(x)}{dt}  +  j_t(x)-j_t(x-1) =0
\label{empiricaldyndir}
\end{eqnarray}

For a given initial condition $N_{t=0}(x)$, 
the measure of Eq. \ref{ptrajmarkovjumpR}
on empirical dynamical trajectory on the time interval $[0,T]$
corresponding to the empirical occupation numbers $N_{0 \leq t \leq T}(.)$ for $x=1,..,L$ and 
to the empirical currents $j_{0 \leq t \leq T}(.)$ related by the empirical dynamics $C^R_{x,t}$ of Eq. \ref{empiricaldyndir}
reads
\begin{eqnarray}
 \int_{C^R_{x,t}} {\cal D} N_t(x) {\cal D} j_t(x) && e^{ -  \displaystyle 
\int_{0}^{T}dt 
\sum_{x=0 }^L  \left[ j_t(x)  \ln \left( \frac{ j_t(x) }{ J_t[x;N_t(.) ]} \right) - j_t(x)   +  J_t[x;N_t(.)] \right]
  }
\label{dirtime}
\end{eqnarray}

\subsection{ Measure for stationary occupations and currents during $[0,T]$  }

When the dynamical rules of Eqs \ref{QJ} and \ref{Qout} do not depend explicitly on the time $t$,
it is natural \cite{maes_onandbeyond,c_open} to consider the probability to observe some stationary occupations $N_t(x)=N(x)$
and some stationary currents $j_t(x)=j(x)$ during some time interval $[0,T]$. 
The empirical dynamics of Eq. \ref{empiricaldyndir}
then reduces to the constraint that the stationary current $j(x)$ should take the same value $j$ on each bond
\begin{eqnarray}
 j(x-1)=j(x) = j
\label{jconst}
\end{eqnarray}
As a consequence, the dynamical measure of Eq. \ref{dirtime} then reduces to the stationary measure
\begin{eqnarray}
 \left[ \prod_{x=1}^L  \int_0^{+\infty}  d N(x) \right] \int_0^{+\infty} dj && e^{ -  T \displaystyle 
\sum_{x=0 }^L  \left[ j  \ln \left( \frac{ j }{ J[x;N(.) ]} \right) - j   +  J[x;N(.)] \right]
  }
\label{dirstatio}
\end{eqnarray}

\subsection{ Statistics of stationary occupations $N(.)$ alone  }

To obtain the measure for the stationary occupations $N(.) $ alone,
one needs to optimize the action of Eq. \ref{dirstatio} over the current $j$
\begin{eqnarray}
0 = \frac{ \partial  } {\partial j } \left( \sum_{x=0 }^L  \left[ j  \ln \left( \frac{ j }{ J[x;N(.) ]} \right) - j   +  J[x;N(.)] \right] \right)
=( L+1)  \ln j - \sum_{x=0 }^{L}  \ln \left(   J[x;N(.) ] \right)   
\label{optimizej}
\end{eqnarray}
Plugging the corresponding optimal current
\begin{eqnarray}
   j^{opt}  [ N(.)  ]  = e^{ \displaystyle \frac{1}{L+1}  \sum_{x=0}^{L} \ln \left( J[x;N(.)]  \right) }
\label{pstatiomarkovjumptreeb1}
\end{eqnarray}
into Eq. \ref{dirstatio} yields 
 the stationary measure for the occupations $N(.)$ alone
\begin{eqnarray}
 \left[ \prod_{x=1}^L  \int_0^{+\infty}  d N(x) \right]  && e^{ -  T \displaystyle 
\left[ 
\sum_{x=0 }^{L} J[x;N(.)]  
- (L +1)  e^{ \displaystyle \frac{1}{L+1}  \sum_{x=0}^{L}  \ln \left( J[x;N(.)]  \right)  }
\right]
  }
\label{dirstatioocc}
\end{eqnarray}
This result is thus the direct generalization of the analogous result obtained for independent particles \cite{c_open}
(see also \cite{c_ring} for the corresponding case of a ring without reservoirs),
the only difference being in the dependence on the constitutive typical values $ J[x;N(.)]  $ on the occupations $N(.)$,
as explained in section \ref{sec_goal}.

\subsection{ Statistics of the stationary current $j$ alone  }

To obtain the measure for the stationary current $j$ alone,
one needs to optimize the action of Eq. \ref{dirstatio} over the occupations $N(y)$ of the system sites $y=1,2,..,L$
\begin{eqnarray}
0  = \frac{ \partial  } {\partial N(y)  }  \left( \sum_{x=0 }^L  \left[ j  \ln \left( \frac{ j }{ J[x;N(.) ]} \right) - j   +  J[x;N(.)] \right] \right)
=\sum_{x=0 }^L \left( 1 - \frac{j} { J[x;N(.)]  }  \right)  \frac{ \partial J[x;N(.)] } {\partial N(y)  } 
\label{deriny}
\end{eqnarray}

So here to go further, one needs to state what precise interactions are present in $ J[x;N(.)] $.

\subsubsection{ Case of zero-range interactions  }

As mentioned in section \ref{sec_goal}, the simplest interactions are the so-called
zero-range interactions where the constitutive typical value $ J[x;N(.)] $ only 
involves the occupation $N_t(x)$ of the initial site $x$ of the jump (Eq. \ref{Qzr})
\begin{eqnarray}
  J^{ZR}[x;N(.)]  = J[x;N(x)] 
\label{zerorange}
\end{eqnarray}
Then the optimization of Eq. \ref{deriny} reduces to the single term
\begin{eqnarray}
0  = \left( 1 - \frac{j} { J[x;N(x)]  }  \right)  \frac{ \partial J[x;N(x)] } {\partial N(x)  } 
\label{derinysingle}
\end{eqnarray}
and we are not interested in cases where $J[x;N(x)] $ does not depend on $N(x)$.
For each site $x=1,2,..,L$ of the system, the optimal occupation $N^{opt}(x)$ is thus determined by the equation
\begin{eqnarray}
 J[x;N^{opt}(x)] = j
\label{derinopt}
\end{eqnarray}
This means that the occupations $N^{opt}(x) $ in the whole system $x=1,2,..,L$
are able to adapt themselves in order to make $ J[x;N^{opt}(x)] $ coincide with the current $j$.
As a consequence, the only remaining contribution in the Lagrangian of Eq. \ref{dirstatio}
comes from the single link between the reservoir $r=0$ and the site $x=1$ of the system,
i.e.  the measure for the current $j$ alone only involves the Poisson Lagrangian
with respect to the external incoming constitutive typical current $ J[0;N(.)] $ from the reservoir $r=0$
\begin{eqnarray}
\int_0^{+\infty} dj   e^{ -  T \displaystyle 
 \left[ j  \ln \left( \frac{ j }{ J[0;N(.) ]} \right) - j   +  J[0;N(.)] \right]
  }
\label{dirstatiozrbord}
\end{eqnarray}

For instance, the case of the power-law dependence of Eq. \ref{Qzrg}
\begin{eqnarray}
 J[x,N(x)]=  \lambda(x) [ N(x)]^g
\label{Jzrg}
\end{eqnarray}
would corresponds to the optimal occupations (Eq \ref{derinopt})
\begin{eqnarray}
N^{opt}(x) = \left( \frac{j}{\lambda(x) }\right)^{\frac{1}{g} } 
\label{noptpower}
\end{eqnarray}

\subsubsection{ Case where $J[x;N(.)]  $ depends on the occupations of the initial site and the final site  }

As mentioned in \ref{sec_goal},
 the next simplest case is when the constitutive typical value $J[x;N(.)] $ only involves the occupations $N(x)$ and $N(x+1)$
of the initial site $x$ and of the final site $(x+1)$ of the jump (Eq. \ref{Qif})
\begin{eqnarray}
 J^{IF}_t[x;N_t(.)] =   J[x;N(x+1);N(x)] 
\label{Jif}
\end{eqnarray}
Then
the optimization of Eq. \ref{deriny} reduces to two terms
\begin{eqnarray}
0  =  \left( 1 - \frac{j} { J[x;N(x);N(x-1)]  }  \right)  \frac{ \partial J[x;N(x);N(x-1)] } {\partial N(x)  } 
+ \left( 1 - \frac{j} { J[x;N(x+1);N(x)]  }  \right)  \frac{ \partial J[x;N(x+1);N(x)] } {\partial N(x)  } 
\label{derinyboth}
\end{eqnarray}
that involves the three consecutive occupations $N(x-1)$, $N(x)$ and $N(x+1)$.
So here the occupation $N(x)$ of the system sites $x=1,2,..,L$
will generically not be able to adapt themselves in order to make $ J[x;N^{opt}(x+1);N^{opt}(x)] $ coincide with the current $j$.
As a consequence, the Lagrangian of Eq. \ref{dirstatio} will have contributions from the whole system $x=1,2,..,L$
in contrast to the zero-range case discussed above (Eq. \ref{dirstatiozrbord}).

\section{ Generalization to other interacting Poisson processes }

\label{sec_compound}

In this paper, we have considered that 

(i) the occupations $N_t(x)$ were integer numbers of particles

(ii) each effective elementary transfer involved a single particle
\begin{eqnarray}
N_t(x) \to N_t(x) -1
\nonumber \\
N_t(y) \to N_t(y) +1
\label{single}
\end{eqnarray}
even if particles interact via the constitutive typical flow $Q_t[y,x;N_t(.)] $ from $x$ to $y$ at time $t$
produced by the occupation $N(.)$ that was assumed to be different
(Eq. \ref{Qdep}) from the independent case of Eq. \ref{Qindep}.

In this section, we briefly mention some possible generalizations within the Poissonian framework.

\subsection{ Allowing the joint-transfer of several particles }

Here we keep the assumption (i) that the occupations $N_t(x)$ are integer numbers of particles,
but we allow the joint-transfer of $n=1,2,..N_t(x)$ particles instead of a single one $n=1$ (Eq \ref{single})
\begin{eqnarray}
N_t(x) \to N_t(x) -n
\nonumber \\
N_t(y) \to N_t(y) +n
\label{several}
\end{eqnarray}
that will take place with the constitutive typical flow $Q_t[y,x;N_t(.);n]  $ that now also depends on 
the number $n$ of particles that jump together. This case can be considered as a special 
quantized version of the next more relevant case involving continuous variables.

\subsection{ Case where the conserved quantity is a continuous variable like the energy }

Here, the occupation $N_t(x)$ that were integer are replaced by continuous variables $E_t(x)$
that will be called energies.
The energy transfer $\omega$ from $x$ to $y$ at time t
is then also a continuous variable $\omega \geq 0$ 
\begin{eqnarray}
E_t(x) \to E_t(x) - \omega
\nonumber \\
E_t(y) \to E_t(y) + \omega
\label{omega}
\end{eqnarray}
and this transfer will  take places with some constitutive typical flow 
$Q_t[y,x;E_t(.); \omega]$.
This means that the generating function of 
Eq \ref{gene} is replaced by
\begin{eqnarray}
< e^{ \nu q_t[y,x] dt } > &&
 = e^{ dt \int_0^{+\infty} d \omega Q_t[y,x;E_t(.); \omega] \left( e^{ \nu \omega }-1 \right) }
\label{genec}
\end{eqnarray}
The Legendre transformation of Eq \ref{legendre} becomes with simplified notations
\begin{eqnarray}
\nu q - {\cal L} (q) && =\int_0^{+\infty} d \omega Q[\omega] \left( e^{ \nu \omega }-1 \right) 
\nonumber \\
\partial_q \left( \nu q -  {\cal L} (q) \right) && =0
\label{legendrec}
\end{eqnarray}
so that the inversion of Eq. \ref{legendreinverse} is replaced by
\begin{eqnarray}
 {\cal L} (q) && = \nu q -  \int_0^{+\infty} d \omega Q[\omega] \left( e^{ \nu \omega }-1 \right) 
\nonumber \\
0 && = \partial_{\nu} \left( \nu q -  \int_0^{+\infty} d \omega Q[\omega] \left( e^{ \nu \omega }-1 \right)    \right) 
= q - \int_0^{+\infty} d \omega Q[\omega] \omega  e^{ \nu \omega }
\label{legendreinversec}
\end{eqnarray}
So depending on the choice of $Q_t[y,x;E_t(.); \omega] $,
it will be possible to compute explicitly $\nu(q)$ and the Lagrangian $ {\cal L} (q)  $,
or it will not be possible to go beyond the parametric form of Eq. \ref{legendreinversec}.
Further work is needed to identify physical interesting models where the Lagrangian ${\cal L} (q)  $ is explicit.

\section{ Conclusions }

\label{sec_conclusion}

In this paper, we have explained how the microscopic Fluctuation Theory (mFT) 
introduced in Ref. \cite{maes_onandbeyond} for a closed system of independent Markov jump processes
and recently extended to the case of open systems with reservoirs \cite{c_open}
could be further generalized to the presence of interactions
that could be either zero-range or involve more neighbors.
We have described the space-time local Lagrangian either in terms of the flows or in terms of the currents.
For the directed version of one-dimensional interacting models between two reservoirs,
we have discussed the consequences and compare with the analogous results obtained previously for independent particles.
Finally, we have briefly mentioned how this approach could be generalized to other types
of interacting Poisson processes involving continuous variables like the energies,
instead of integer numbers of particles.

%%%%%%%%%%%%%%%%%%%%%%%%%%%%%%%%%%%%%%%%%%%%%%%%%%%%

\end{document}